\documentclass[12pt]{article}
\usepackage[utf8]{inputenc}
\usepackage{amsmath, amssymb}
\usepackage{graphicx}
\usepackage{caption}
\usepackage{geometry}
\usepackage{authblk}
\usepackage{cite}
\usepackage{hyperref}
\usepackage{subcaption}
\usepackage{float}
\usepackage{xcolor}

\title{Intermediate Thermal Equilibrium Stages in Molecular Dynamics Simulations of two Bodies in Contact}

\author[1,3]{Jonathas N. da Silva}
\author[2]{Octavio D. Rodriguez Salmon}
\author[2] { Minos A. Neto}

\affil[1]{Fundação Centro de Análise Pesquisa e Inovação Tecnológica. , Av. Gov. Danilo de Matos Areosa, 381 - Distrito Industrial I, Manaus - AM, 69075-351 - Brasil}
\affil[2]{Departamento de Física, Universidade Federal Do Amazonas, 3000, Japiim, 69077-000, Manaus, AM, Brazil}
\affil[3]{Laboratório de Química Teórica e Computacional, Universidade Federal Do Amazonas, 3000, Japiim, 69077-000, Manaus, AM, Brazil}

\date{}

\begin{document}

\maketitle

\begin{abstract}
The Zeroth Law of Thermodynamics states that if two systems are in thermal equilibrium with a third one, then they are also in equilibrium with each other. This study explores not only the final state of thermal equilibrium between ideal gases separated by heat-conducting walls, but also the intermediate stages leading up to equilibrium, using classical molecular dynamics simulations. Two- and three-region models with argon atoms are analyzed. Fluctuations, correlations, and temperature distributions are observed, highlighting how heat conduction between regions influences the time to reach equilibrium. This work is distinguished by its detailed analysis of the intermediate stages that occur until the system reaches thermal equilibrium, in accordance with the Zeroth Law of Thermodynamics.
\end{abstract}

\section{Introduction}

In complex thermodynamic systems, the presence of local and global maxima defines the coexistence of metastable states and absolute thermodynamic equilibrium, where the free energy landscape governs system stability against stochastic fluctuations. This multiplicity of extremes is crucial in non-equilibrium thermodynamics for dissipative structures and non-equilibrium steady states (NESS), where open systems can be trapped in local maxima of dynamic stability that optimize energy dissipation. This phenomenon is observed in cellular biophysics, where chemical gradient control prevents global equilibrium, and in condensed matter physics, where glass transitions show trapping in local energy minima due to kinetic barriers. The evolution of systems far from equilibrium is often ruled by extremality principles, such as Maximum Entropy Production (MaxEP), which explain the statistical preference for self-organizing configurations representing local peaks of dissipative efficiency over thermal rest. Recent global frameworks for stationary non-equilibrium states have emphasized
the role of state parameters beyond equilibrium descriptions.~\cite{Holyst2025}

The Zeroth Law of Thermodynamics is fundamental to the definition of temperature and thermal equilibrium. In essence, it states that if two systems are in thermal equilibrium with a third one, then they are in equilibrium with each other. While this principle is widely used in thermodynamics, it rarely addresses the dynamic process that leads to equilibrium, and numerical studies have even reported apparent inconsistencies in
thermostatted settings.~\cite{PatraBhattacharya2018}. \\

From a foundational viewpoint, thermodynamic laws are classically formulated for equilibrium states, where entropy and state functions provide a complete macroscopic description, while the microscopic dynamical route toward equilibrium
is not addressed within the axiomatic equilibrium framework.~\cite{LiebYngvason1999}

This study employs molecular dynamics simulations to investigate the details of the equilibration of two bodies or systems in thermal contact. To explore the equilibration process, we adopted the classical Molecular Dynamics (MD) approach, which allows the direct simulation of atomic trajectories through the numerical integration of Newton's equations of motion. Interatomic interactions are modeled via the Lennard-Jones potential, which is a well-established choice for noble gasses like argon.  

Accordingly, ideal gases of argon atoms were simulated in two different configurations, separated by thin walls that allow heat transfer. The analysis includes the temporal evolution of temperature and the mechanisms of heat transfer, with a focus on the dynamic behavior of fluctuations, correlations, and temperature distributions until thermal equilibrium is reached.


\section{Methodology}

Molecular Dynamics (MD) is a computational technique that models the motion of atoms and molecules on the basis of classical mechanics. The fundamental assumption is that each particle obeys Newton’s second law:
\begin{equation}
m_i \frac{d^2 \vec{r}_i}{dt^2} = \vec{F}_i,
\end{equation}
where $m_i$ is the mass of the particle $i$, $\vec{r}_i$ is its position vector, and $\vec{F}_i$ is the total force acting on it.

The force $\vec{F}_i$ is computed as the negative gradient of the interatomic potential $U$, typically modeled by the Lennard-Jones (LJ) potential for noble gases:
\begin{equation}
  U(r) = 4\varepsilon \left[ \left( \frac{\sigma}{r} \right)^{12} - \left( \frac{\sigma}{r} \right)^6 \right],
\end{equation}
where $r$ is the distance between the atoms, $\varepsilon$ is the depth of the potential well, and $\sigma$ is the finite distance at which the potential between particles is zero. 

By integrating the equations of motion over time, MD simulations produce trajectories for all particles, from which macroscopic thermodynamic quantities such as temperature and pressure can be calculated. Integration is typically performed using algorithms such as the velocity-Verlet method, which provides good numerical stability and energy conservation properties.
\color{black}

\begin{figure}[H]
    \centering
    \begin{subfigure}{6cm}
        \centering
        \includegraphics[width=5.6cm]{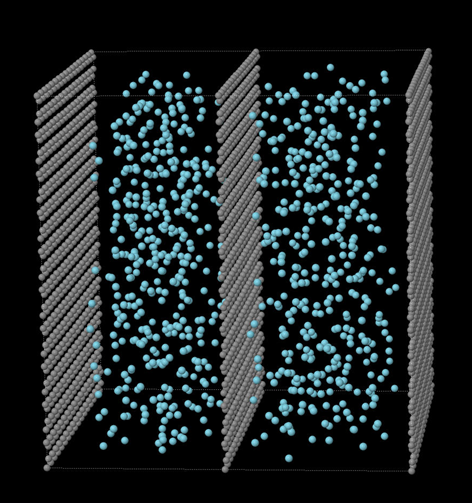}
        \caption{}
        \label{fig:case1}
    \end{subfigure}
    \hspace{1cm}
    \begin{subfigure}{6cm}
        \centering
        \includegraphics[width=6cm]{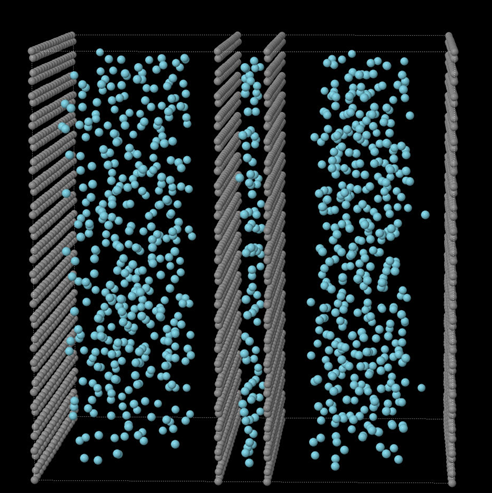}
        \caption{}
        \label{fig:case2}
    \end{subfigure}
    \caption{(a) Simulation box with three walls separating the system into two regions (left and right), each containing 400 argon atoms. (b) Simulation box with four walls separating the system into three regions (left, middle, and right), with 400 argon atoms in the lateral regions and 100 atoms in the central region.}

        \label{fig:cases}
\end{figure}

The molecular dynamics simulation~\cite{frenkel2001} was performed in a box of 40\,Å $\times$ 40\,Å $\times$ 40\,Å containing argon atoms, modeled using the Lennard-Jones potential~\cite{allen1989}. The separating walls were simulated as graphene sheets, which also interact via the Lennard-Jones potential. The system was set up for two cases:

\begin{itemize}
    \item \textbf{Case 1:} Two regions (left and right) containing 400 argon atoms each, separated by three thin walls (Figure~\ref{fig:case1}). The initial temperatures were set to 100\,K and 500\,K for the left and right regions, respectively.
    
    \item \textbf{Case 2:} Three regions (left, middle, and right), with 400 atoms in the lateral regions and 100 atoms in the central region, with temperatures of 100\,K, 300\,K, and 500\,K, respectively (Figure~\ref{fig:case2}). The central region has a volume eight times smaller than the lateral regions.
\end{itemize}

Initially, each region was equilibrated using the NVT ensemble (constant Number of particles, Volume, and Temperature). After this stage, the simulations proceeded with the NVE ensemble (constant Number of particles, Volume, and Energy), allowing for the analysis of thermal transfers between the regions.

Classical molecular dynamics simulations were performed using the \textit{LAMMPS} (Large-scale Atomic/Molecular Massively Parallel Simulator) package~\cite{plimpton1995fast}.

The simulations were conducted with \texttt{real} units, appropriate for representing atomic-scale systems with values directly related to physical quantities such as temperature in kelvins and time in units of multiples of seconds. The atomic style was set to \texttt{atomic}, suitable for species without explicit bonds, such as noble gases and fixed solid structures.~\cite{lammps_doc}.

The simulation domain was three-dimensional, with periodic boundary conditions in the $x$ and $y$ directions, and a fixed boundary at $z$, representing a physical barrier that confines the atoms in the direction of heat transfer.

The initial velocities were randomly assigned to the \texttt{left} and \texttt{righ} regions (case 1) to generate a thermal gradient: 100~K on the left and 500~K on the right and \texttt{left}, texttt{midlle} and \texttt{righ} for case 2 with the same conditions at the ends and 300~K in the middle. This arrangement allows simulation of spontaneous heat flow between the two ends in both cases. ~\cite{muller1997simple}.

Interatomic interactions were modeled using the truncated Lennard-Jones potential (\texttt{lj/cut}) with different parameters for each type of interaction:

\begin{itemize}  
  \item Ar--Ar: $\varepsilon = 0{,}25$ kcal/mol, $\sigma = 3{,}3$~Å, cut in 8{,}4~Å  
  \item Ar--C: $\varepsilon = 0{,}12$ kcal/mol, $\sigma = 3{,}4$~Å, cut in 8{,}5~Å  
  \item C--C: $\varepsilon = 0{,}06$ kcal/mol, $\sigma = 3{,}4$~Å, cut in 8{,}5~Å
\end{itemize}

The neighborhood was calculated with a cutoff distance of 2{,}0~Å and was updated every 8 steps. To reduce computational cost, interactions between (immobile) carbon atoms were excluded.

An initial relaxation step was conducted using Langevin dynamics with a damping factor of 100{,}0~fs and a time step of 0{,}1~fs. The objective was to allow thermal and structural relaxation of the mobile atoms (fluid), given that the carbon wall atoms were kept fixed.

The main simulation was conducted in the NVT ensemble (constant particle number, volume, and temperature), with temperature control at 300~K through the Nose--Hoover algorithm, using a thermal coupling time of 100{,}0~fs. Only the mobile atoms were integrated by the equation of motion, and the temperature was computed specifically for these degrees of freedom ~\cite{hoover1985canonical}.

The simulation was run for $10^7$ steps with a time step of 1{,}0~fs, totaling 10~ns of physical time, a common value for observing thermal relaxation processes in atomic systems~\cite{hoover1985canonical}. The simulation output was recorded every 500 steps for later structural and energetic analysis with emphasis on the thermal evolution of the system and the stabilization of the heat flux ~\cite{schelling2002comparison}.

To properly simulate diathermal walls, the interaction energy parameter ($\epsilon$) of the carbon atoms forming the walls was intentionally reduced, allowing the internal walls to spatially separate the subsystems without hindering thermal energy exchange between them. In contrast, the lateral walls interact with atoms only on one side and, due to the absence of periodic boundary conditions in those directions, effectively behave as adiabatic walls. This combination of diathermal and adiabatic walls ensures that heat exchange occurs exclusively among the subsystems, thereby providing controlled physical conditions for observing thermal equilibrium as established by the Zeroth Law of Thermodynamics.
 
\section{Results and Discussions}

The configuration of the walls, with effectively diathermal internal interfaces and adiabatic external walls, ensures that the exchange of thermal energy occurs exclusively between the subsystems considered, without influence from the external environment. This guarantees that the system satisfies the appropriate conditions for observing thermal equilibrium according to the Zeroth Law of Thermodynamics, which formalizes the notion of equivalence relation: if two subsystems are in thermal equilibrium with a third, then they are in thermal equilibrium with each other. In the following sections, we will discuss how, microscopically, the temperatures of the different regions evolve and equalize over time, connecting the observed behaviors to the principle of thermal equivalence.

\subsection{Temperature Evolution and Thermal Equilibrium}

\begin{figure}
    \centering
    \includegraphics[width=12cm]{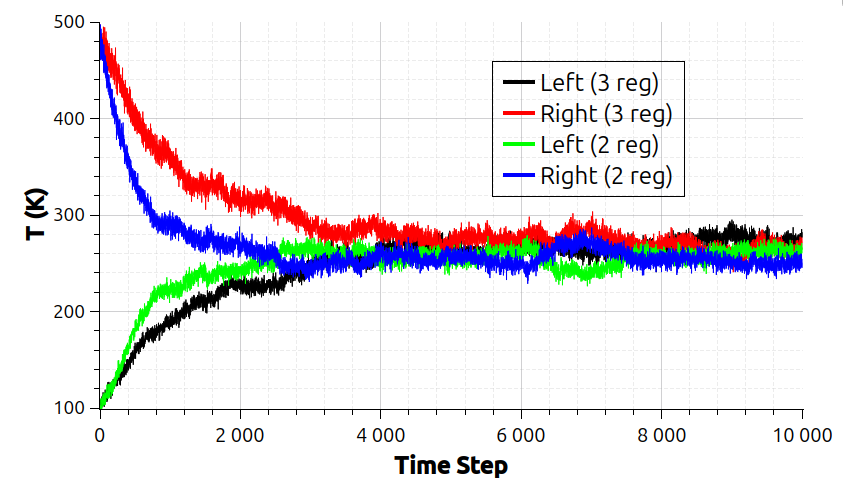}
    \caption{Evolution of temperature for a system with two and three regions in contact. Comparative analysis of temperature evolution in the lateral regions for Cases 1 and 2.}
    \label{fig:fig2}
\end{figure}

\begin{figure}
    \centering
    \includegraphics[width=14cm]{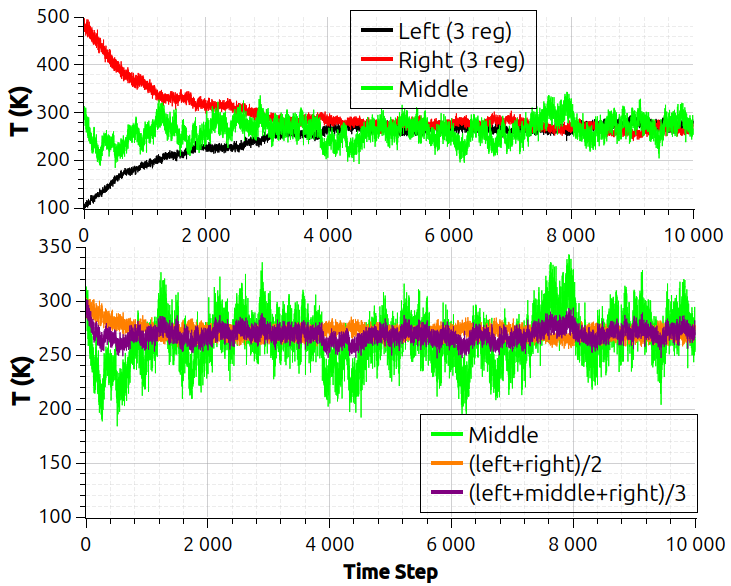}
    \caption{Evolution of temperature for a system with three regions in contact.Temperature profiles for each region in Case 2 (top). The bottom panel presents a comparison between the average temperature of the lateral regions, the global average temperature, and the temperature in the middle region.
}
    \label{fig:fig3}
\end{figure}

\begin{figure}
    \centering
    \includegraphics[width=12cm]{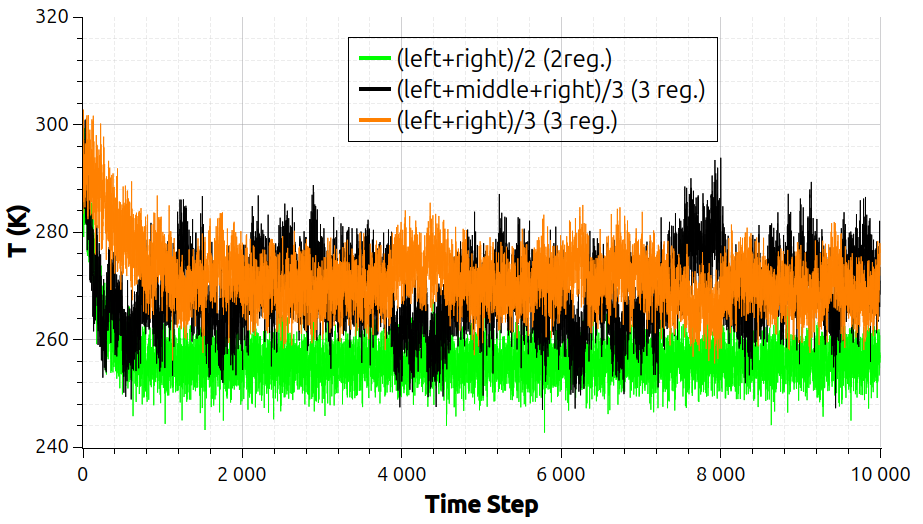}
    \caption {Evolution of the temperature means for cases 1 and 2. Comparative Evolution of the average temperature for a system with two and three regions in contact. }
    \label{fig:fig4}
\end{figure}

The results showed that, in Case 1, the temperatures in the lateral regions converge almost exponentially to an average value, as seen in Figure~\ref{fig:fig2}. The process is relatively straightforward, with no significant fluctuations over time. In contrast, in Case 2, the presence of the middle region leads to more complex behavior, as shown in Figure~\ref{fig:fig3}, where the temperature in the central region exhibits greater variations before reaching an equilibrium state. Figure~\ref{fig:fig4} compares the evolution of the average temperature between the two cases, showing that thermal equilibrium in Case 2 occurs more slowly due to the role of the middle region as a heat transport mediator.

\subsection{Conditional temperature volatility and Distributions}

\begin{figure}
    \centering
    \includegraphics[width=16cm]{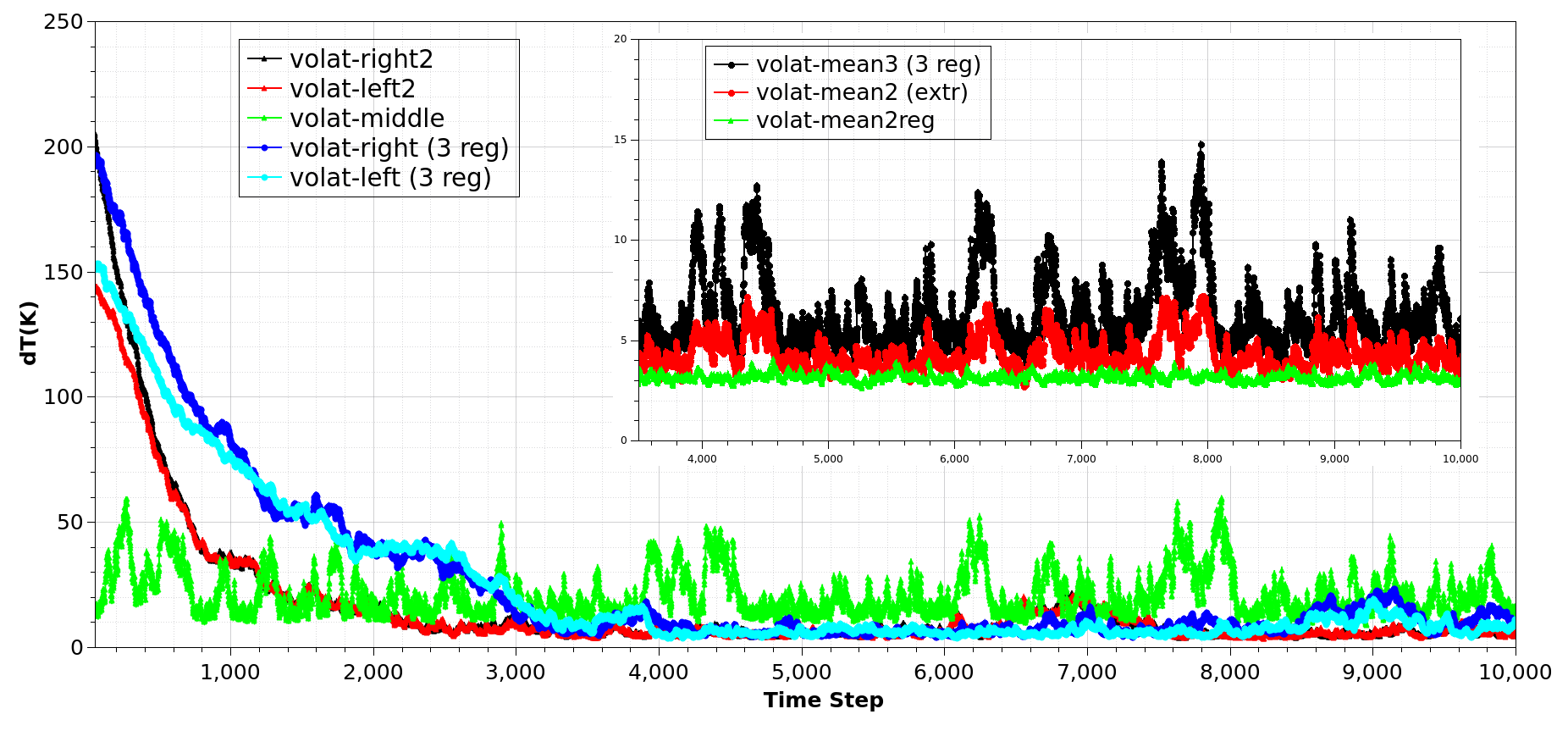}
    \caption {Conditional temperature volatility (dT) for two and three regions in contact. The triangular points correspond to Case 1 and the circular points to Case 2. The subgraph shows the average value for Case 1 (average of the two regions) and Case 2 (average of the ends and the three regions together).}
    \label{fig:fig5}
\end{figure}

\begin{figure}[H]
    \centering
    \begin{subfigure}{10cm}
        \centering
        \includegraphics[width=10cm]{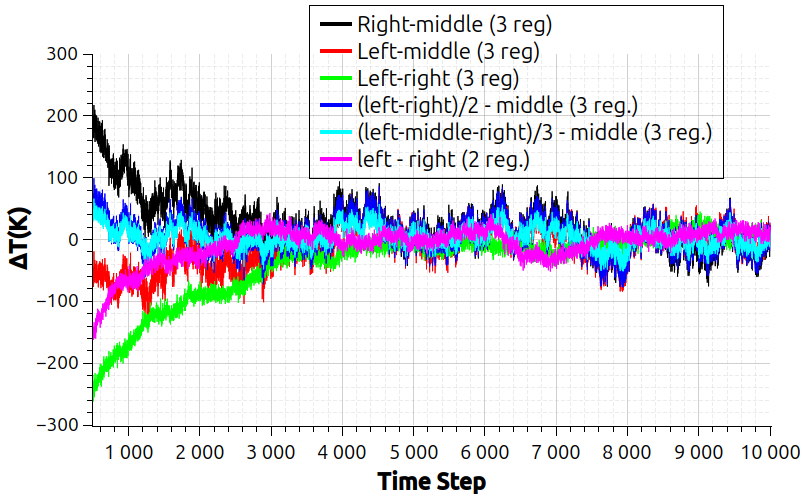}
        \caption{}
        \label{fig:fig6a}
    \end{subfigure}
   
    \begin{subfigure}{10cm}
        \centering
        \includegraphics[width=10cm]{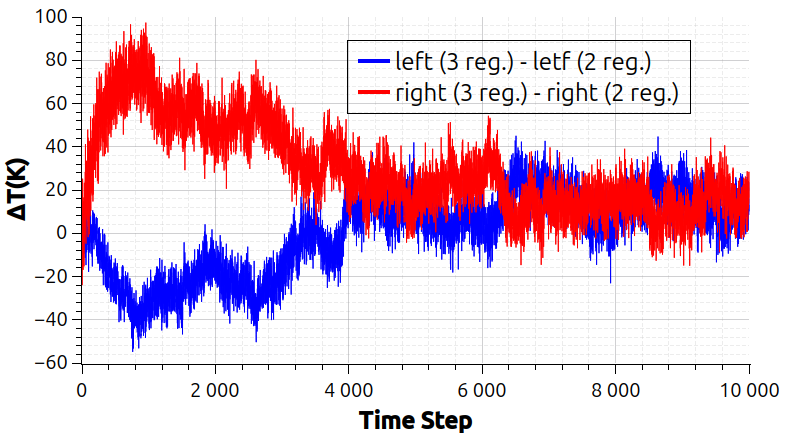}
        \caption{}
        \label{fig:fig6b}
    \end{subfigure}
    \caption{(a) Evolution of difference of temperature for a system with three regions in contact. (b) Difference of temperature between two systems (three regions and two regions), cases 1 and 2.}

        \label{fig:cases}
\end{figure}

For the quantitative characterization of the temporal variability of temperature, the GARCH (Generalized Autoregressive Conditional Heteroscedasticity) model was employed, widely used in the analysis of time series with time-dependent conditional variance ~\cite{Bollerslev,tsallis}. Unlike approaches based solely on global variance, GARCH allows for the description of the dynamics of conditional temperature volatility, capturing distinct statistical regimes of greater or lesser intensity of variations throughout the temporal evolution. This approach is particularly suitable for data from molecular dynamics simulations, in which temperature—defined microscopically from kinetic energy—exhibits time-correlated physical fluctuations due to the continuous exchange of energy between finite subsystems; in this context, GARCH acts as a statistical tool to quantify the persistence and intermittency of thermal variability. Thus, the model makes it possible to robustly identify and compare the different thermal volatility regimes associated with the different regions of the system, as well as statistically discriminate the stages of the thermal relaxation process until equilibrium. Temperature volatility revealed that the lateral regions in Case 2 reach lower intensity, while the middle region experiences larger variations, as illustrated in Figure~\ref{fig:fig5}. Figure~\ref{fig:fig6a} compares the temperature differences between regions in both cases, while Figure~\ref{fig:fig6b} highlights the variations between Cases 1 and 2. \\
Also, as illustrated in Figure~\ref{fig:fig5}, the conditional temperature volatility in the lateral regions progressively decreases until reaching equilibrium around 3500 time steps. In the system composed of only two regions, this regime is reached more quickly, approximately in 2500 time steps. In contrast, the central region exhibits intermittent behavior, characterized by volatility clusters throughout the time evolution, notably in the intervals of [4000, 4500] and [6000, 8000] time steps. This pattern indicates that energy exchange in this region occurs non-uniformly, dominated by localized events of greater intensity. Furthermore, even after the lateral regions reach equilibrium, evidenced by volatility values close to zero, the central region maintains high levels, showing that subsystems with fewer atoms are more susceptible to energy fluctuations, which is reflected in more intense and persistent local thermal variations.

\begin{figure}[H]
    \centering
    \begin{subfigure}{7cm}
        \centering
        \includegraphics[width=9cm]{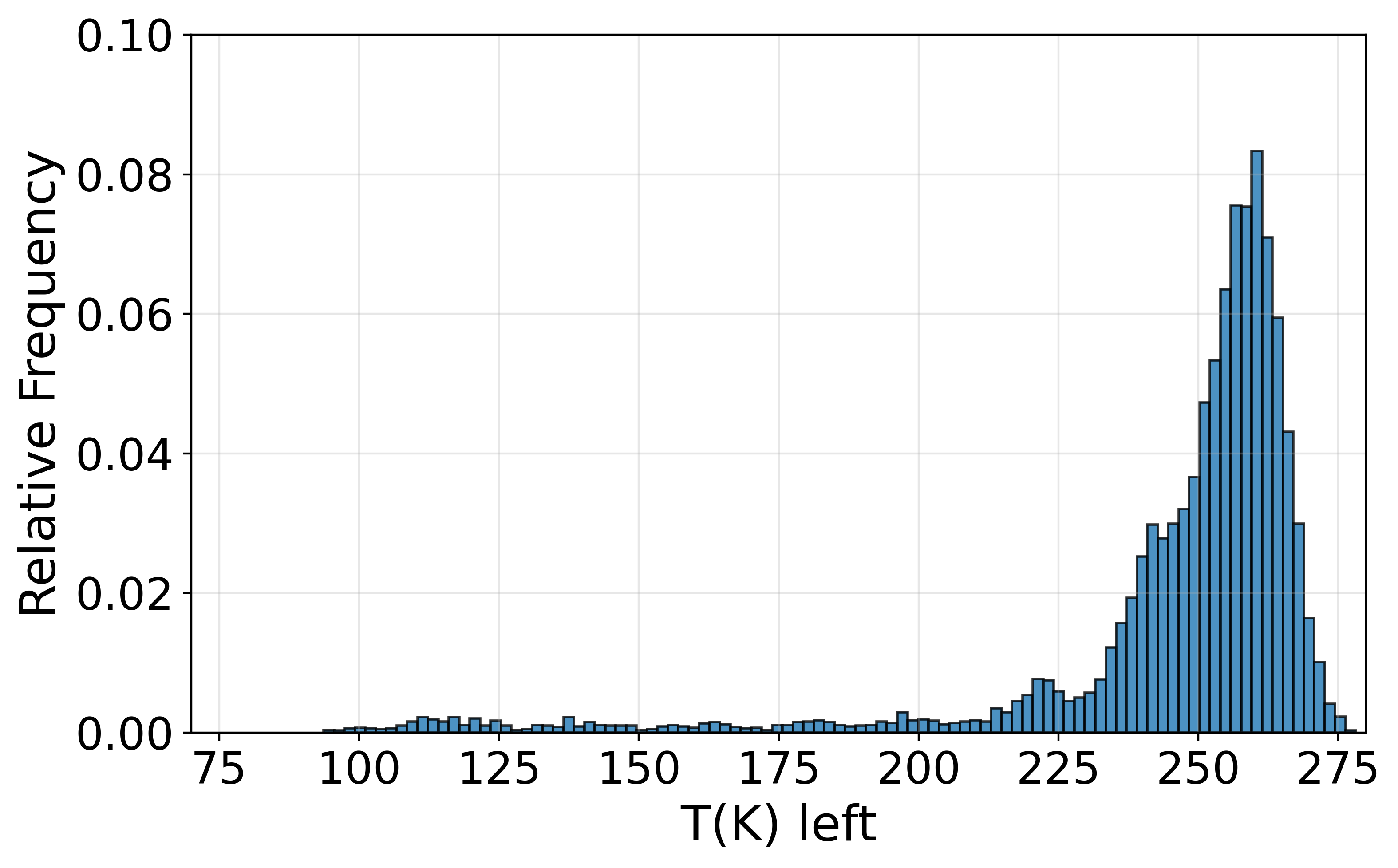}
        \caption{}
        \label{fig:fig7a}
    \end{subfigure}
    \hspace{1.8cm}
    \begin{subfigure}{7cm}
        \centering
        \includegraphics[width=9cm]{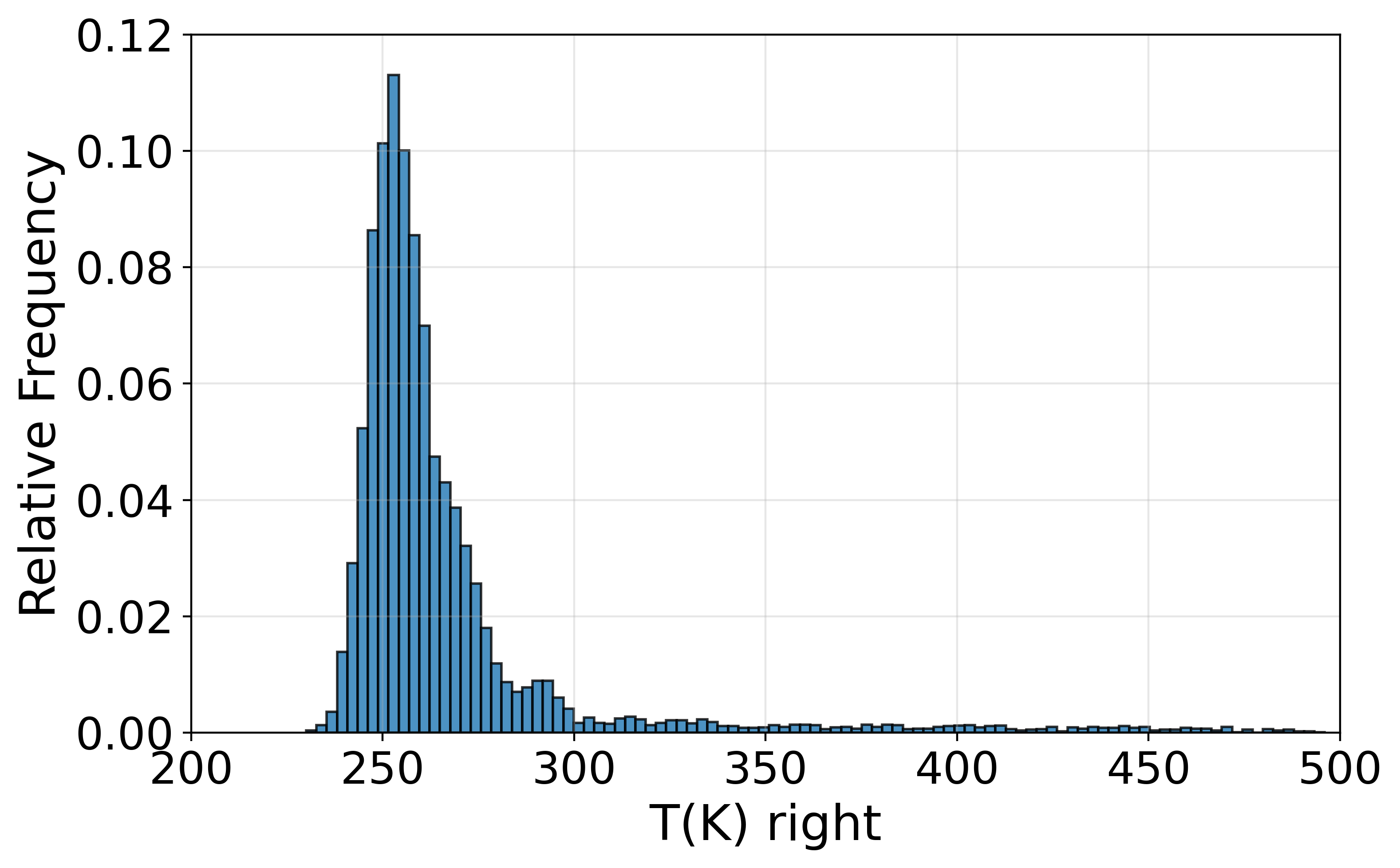}
        \caption{}
        \label{fig:fig7b}
    \end{subfigure}
    \caption{Temperature frequency distribution for the left (a) and right (b) regions in case 1.}
        \label{fig:fig7}
\end{figure}

\begin{figure}[H]
    \centering
    \begin{subfigure}{7cm}
        \centering
        \includegraphics[width=9cm]{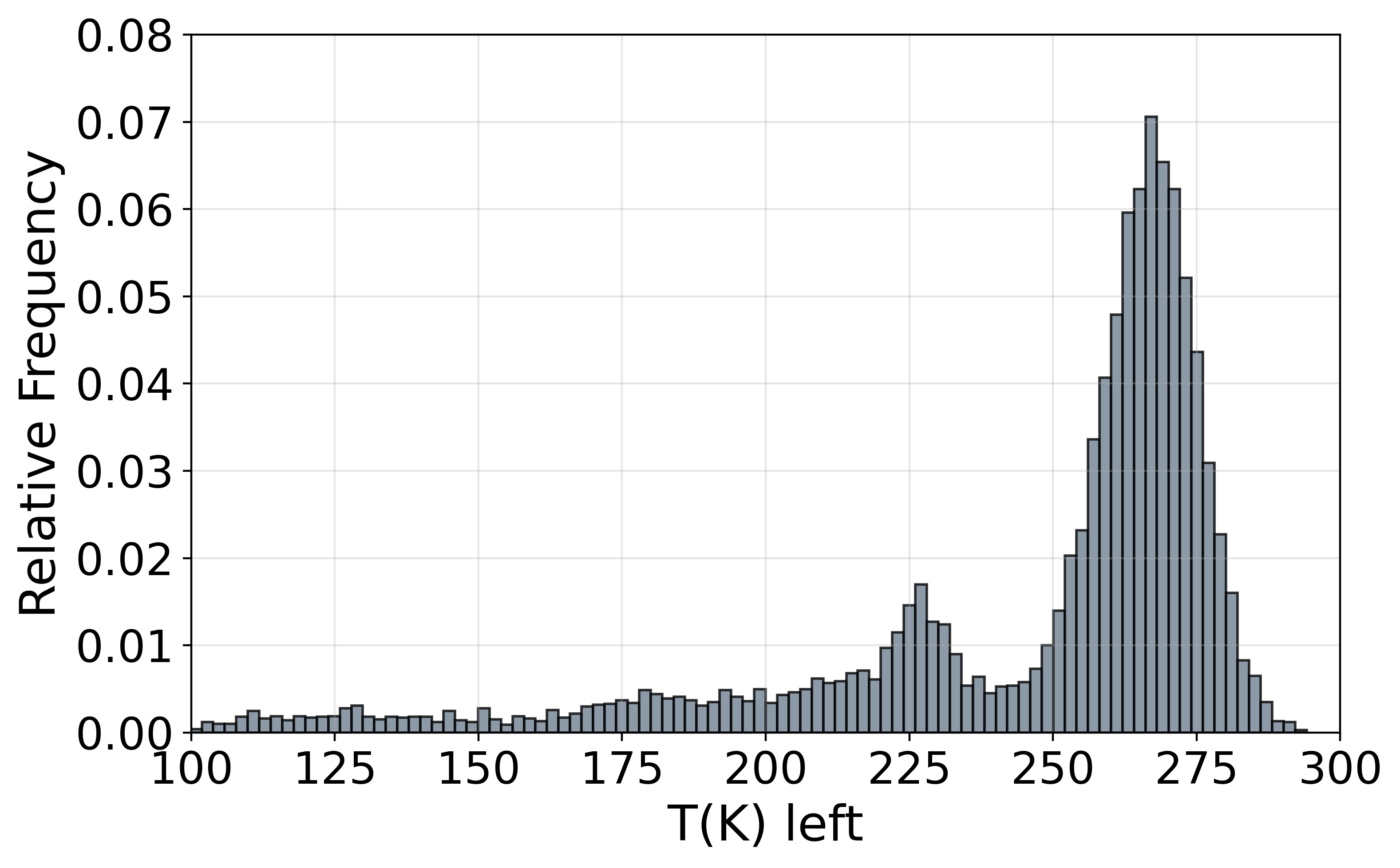}
        \caption{}
        \label{fig:fig8a}
    \end{subfigure}
    \hspace{1.8cm}
    \begin{subfigure}{7cm}
        \centering
        \includegraphics[width=9cm]{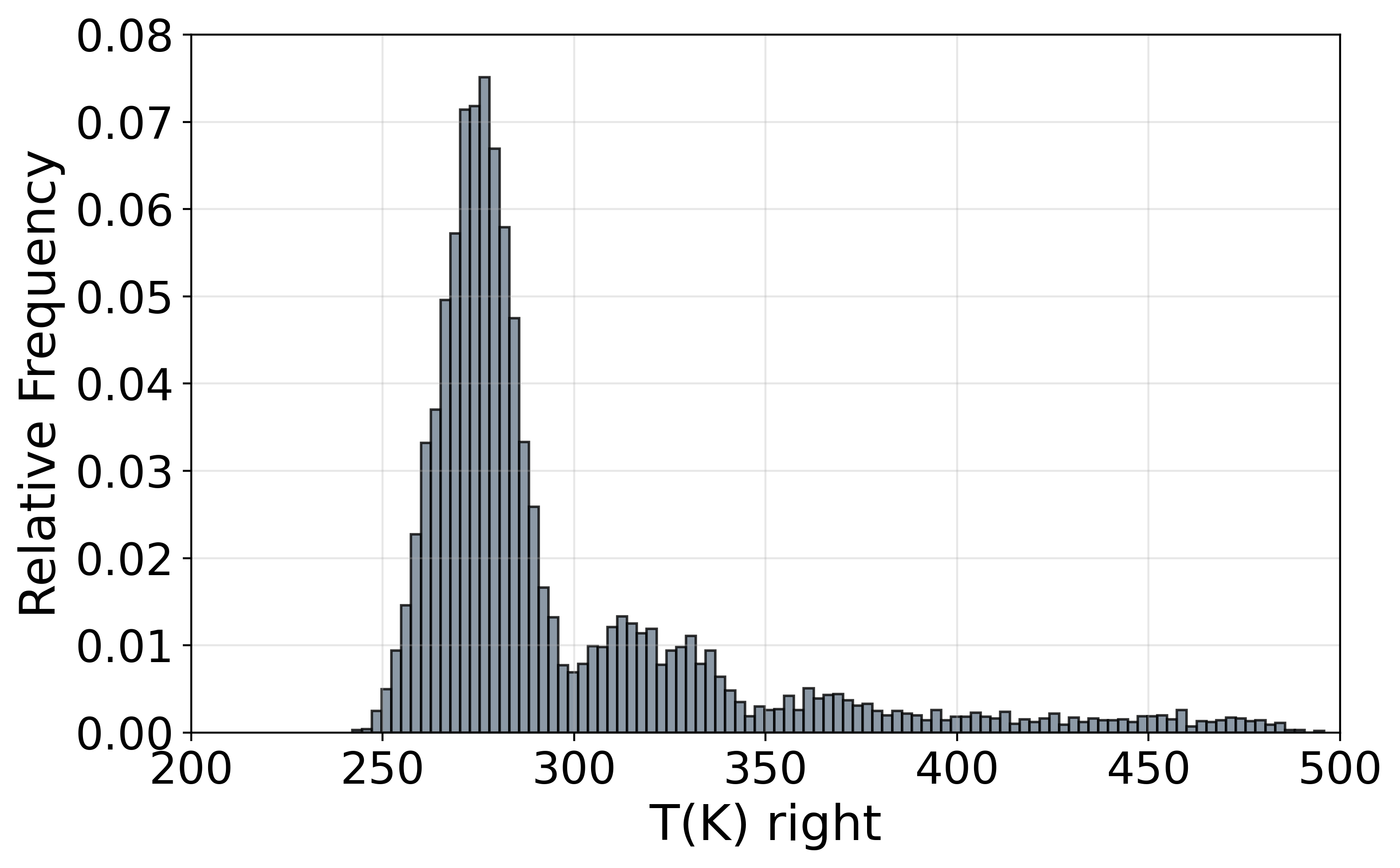}
        \caption{}
        \label{fig:fig8b}
    \end{subfigure}
 \begin{subfigure}{7cm}
        \centering
        \includegraphics[width=9cm]{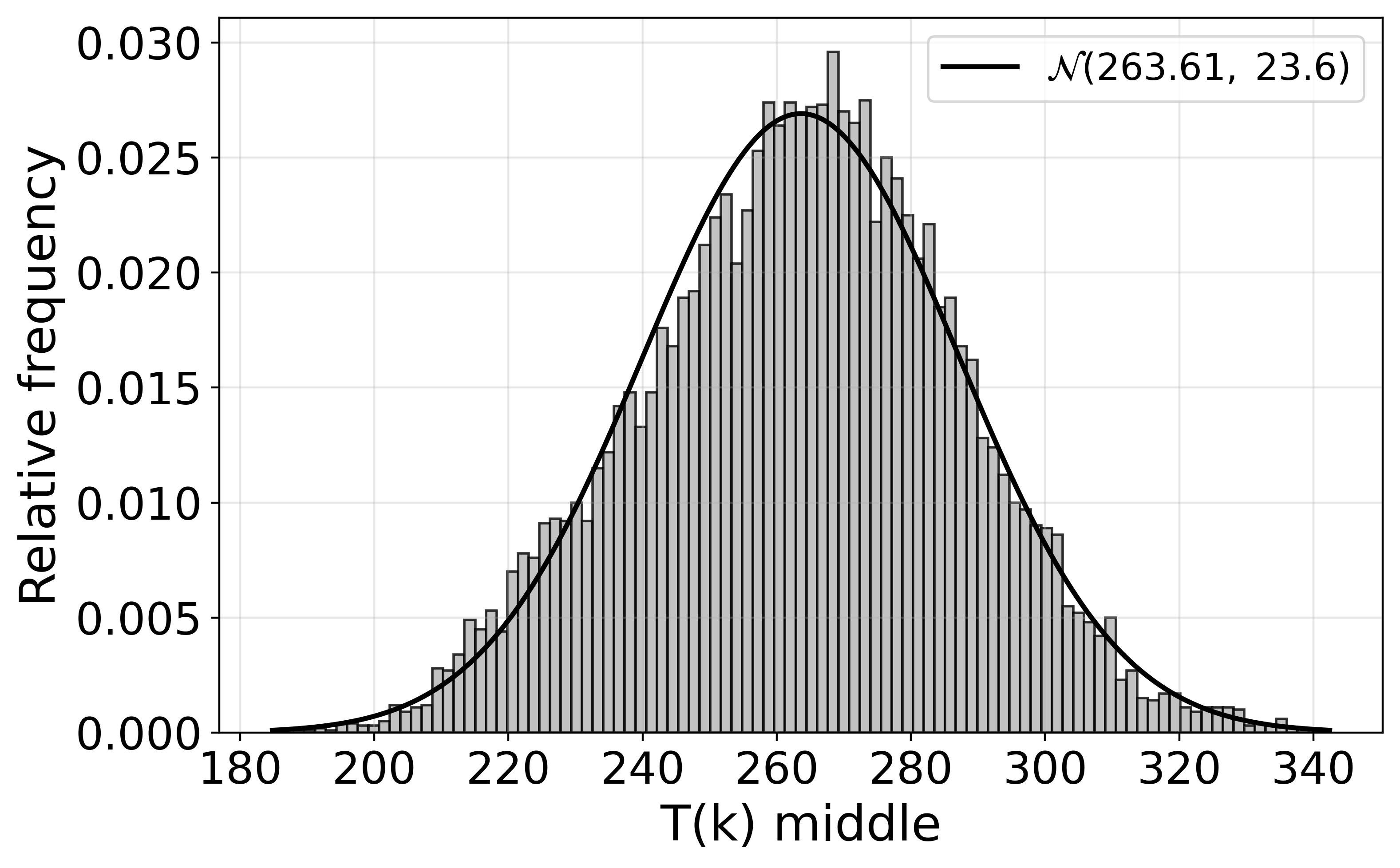}
        \caption{}
        \label{fig:fig8c}
    \end{subfigure}
    
    \caption{Temperature frequency distribution: (a) for the left region; (b) right region, (c) middle region, in case 2.}
        \label{fig:fig8}
\end{figure}

Furthermore, the distribution of temperature frequency (Figures~\ref{fig:fig7} and~\ref{fig:fig8}) shows that, particularly in Case 2 (Figures~\ref{fig:fig8a} and~\ref{fig:fig8b}), the lateral regions display a main peak and a secondary peak, suggesting that different temporary equilibrium states occur before reaching the final equilibrium. Figure~\ref{fig:fig8} presents the temperature distribution for the three regions in Case 2, with the middle region (Figure~\ref{fig:fig8c}) showing a wider spread of values, reflecting the difficulty of achieving thermal equilibrium.


\

The emergence of a bimodal temperature distribution in the three-region model reveals a temporal breakdown of transitivity as dictated by the Zeroth Law of Thermodynamics. Although the system eventually converges to a single global maximum representing the final equilibrium, the intermediate stages exhibit a local maximum that characterizes a metastable state between two adjacent regions. This behavior suggests that thermal synchronization occurs at different time scales, where the thermal resistance at the interfaces prevents the instantaneous homogenization of the kinetic energy across all regions, leading to the coexistence of distinct local equilibrium states ~\cite{PatraBhattacharya2018}.

\subsection{Temperature Differences and Correlations}

\begin{figure}[H]
    \centering
    \begin{subfigure}{7cm}
        \centering
        \includegraphics[width=9cm]{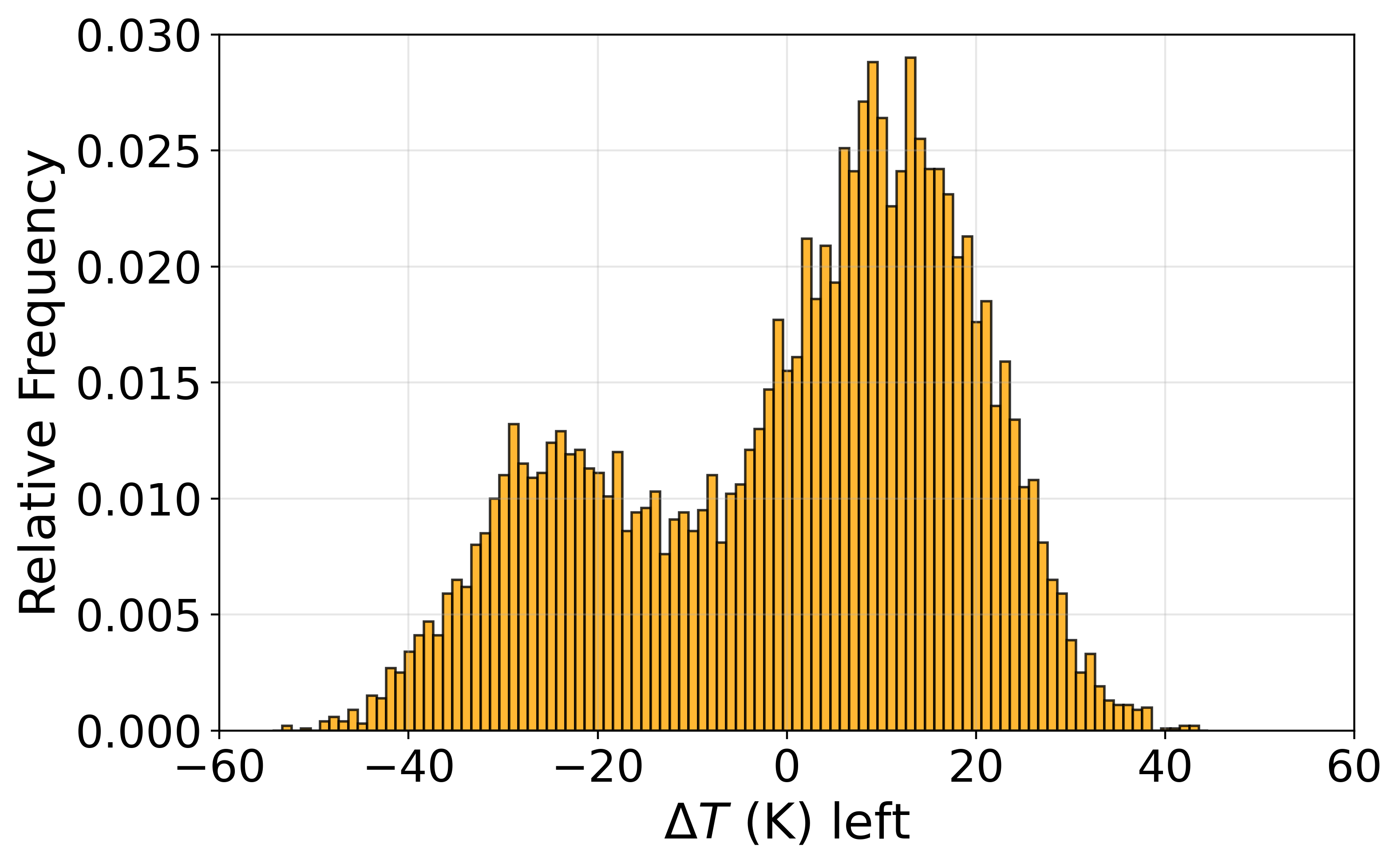}
        \caption{}
        \label{fig:9a}
    \end{subfigure}
    \hspace{1.5cm}
    \begin{subfigure}{7cm}
        \centering
        \includegraphics[width=9cm]{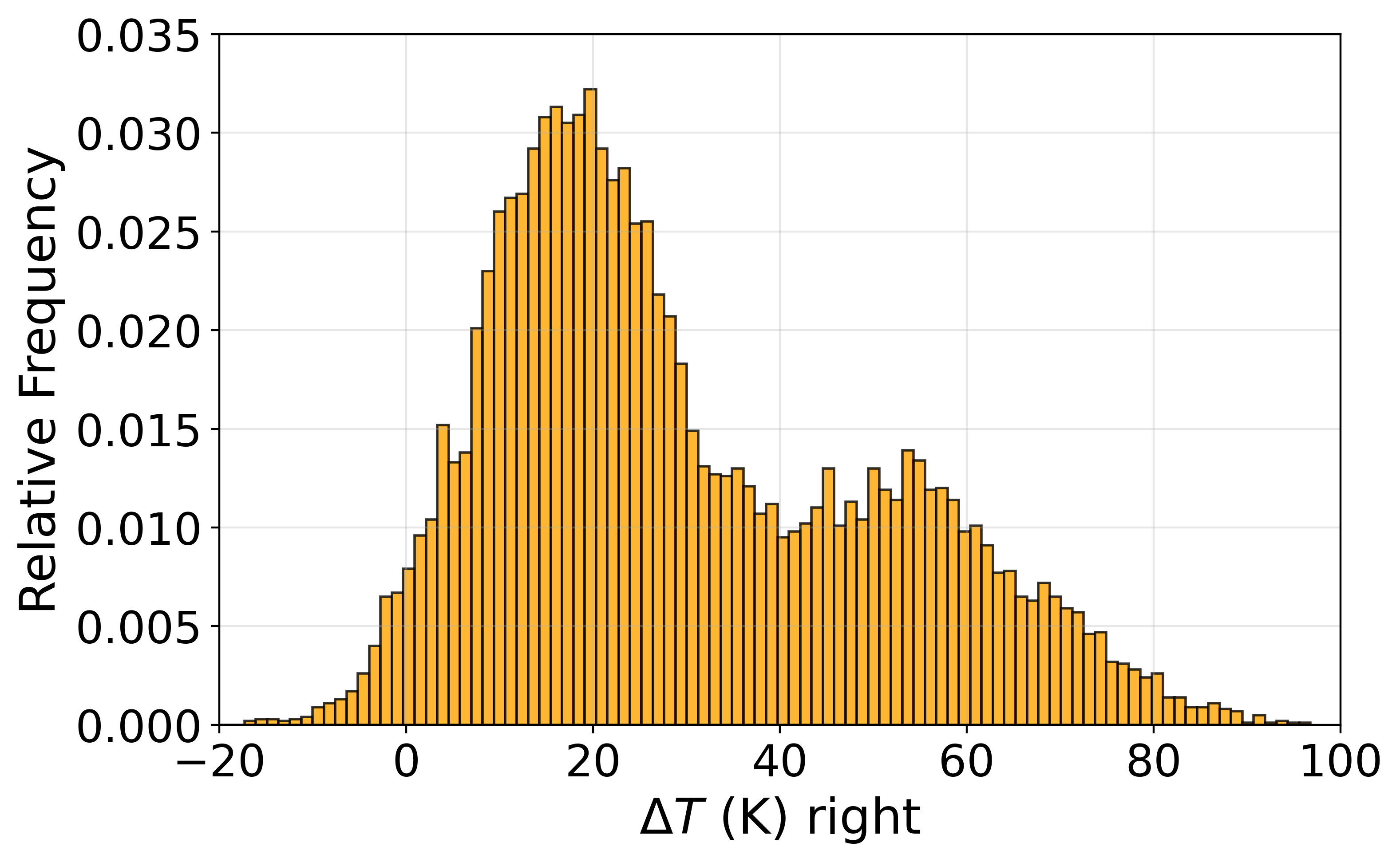}
        \caption{}
        \label{fig:9b}
    \end{subfigure}
    \caption{Frequency distribution for the temperature difference between case 1 and case 2 in the left (a) and right (b) regions.}
        \label{fig:9}
\end{figure}

\begin{figure}[H]
    \centering
    \begin{subfigure}{7cm}
        \centering
        \includegraphics[width=9cm]{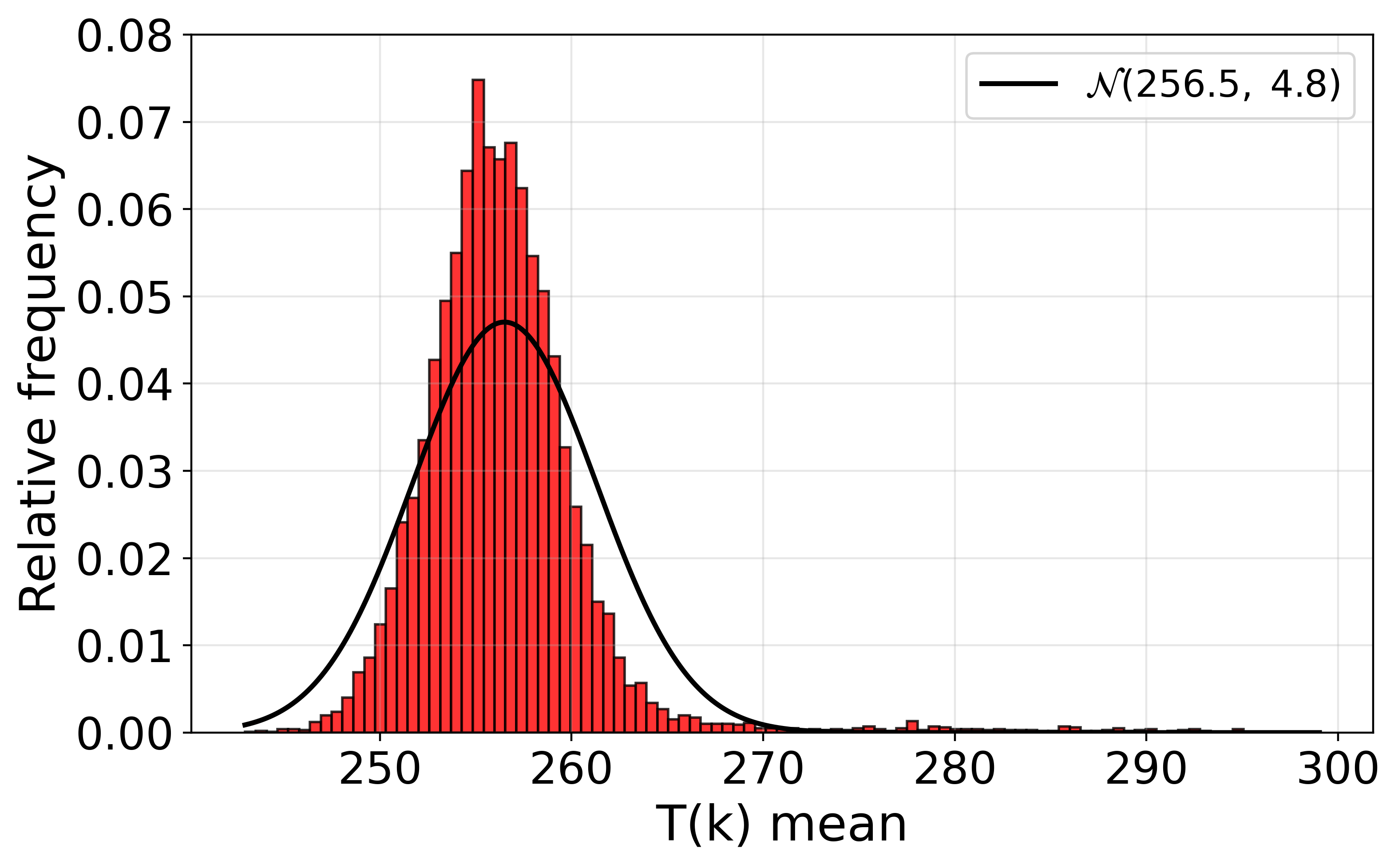}
        \caption{}
        \label{fig:10a}
    \end{subfigure}
    \hspace{1.5cm}
    \begin{subfigure}{7cm}
        \centering
        \includegraphics[width=9cm]{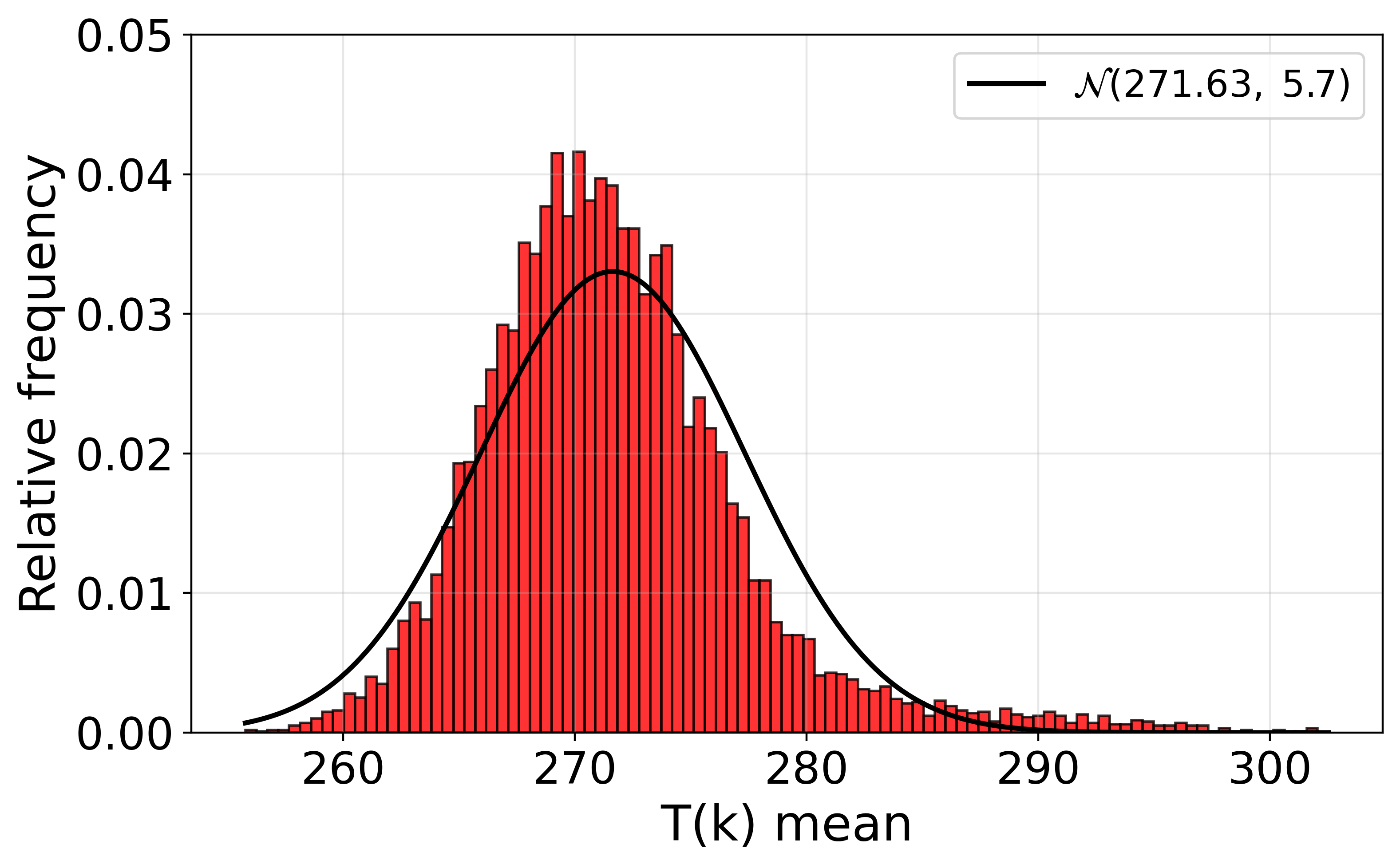}
        \caption{}
        \label{fig:10b}
    \end{subfigure}
    \caption{Frequency distribution for the average system temperature in case 1 (a) and 2 (b).}
        \label{fig:10}
\end{figure}

Figures~\ref{fig:9a} and~\ref{fig:9b} show the frequency distributions of temperature differences between the regions in both cases. It is observed that, in Case 2, the average temperature difference is greater and more unstable, reflecting the additional fluctuations in the middle region. Figures~\ref{fig:10a} and~\ref{fig:10b} illustrate the distribution of the systems' average temperature, revealing a leptokurtic behavior, especially in Case 1, which shows a stronger concentration of temperatures around specific values. \\

\begin{figure}[H]
    \centering
    \begin{subfigure}{7cm}
        \centering
        \includegraphics[width=9cm]{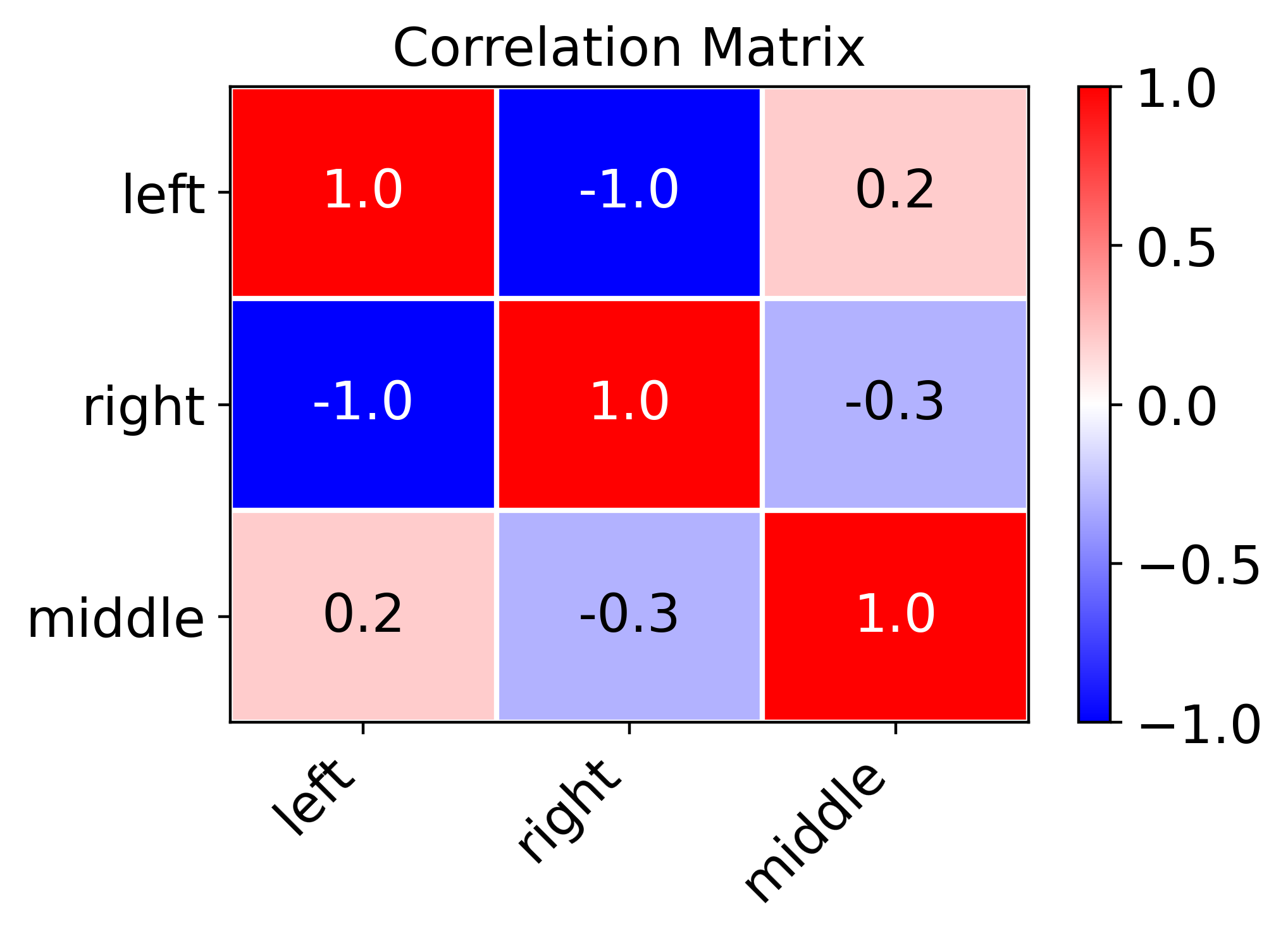}
        \caption{}
        \label{fig:fig11a}
    \end{subfigure}
    \hspace{1.8cm}
    \begin{subfigure}{7cm}
        \centering
        \includegraphics[width=9cm]{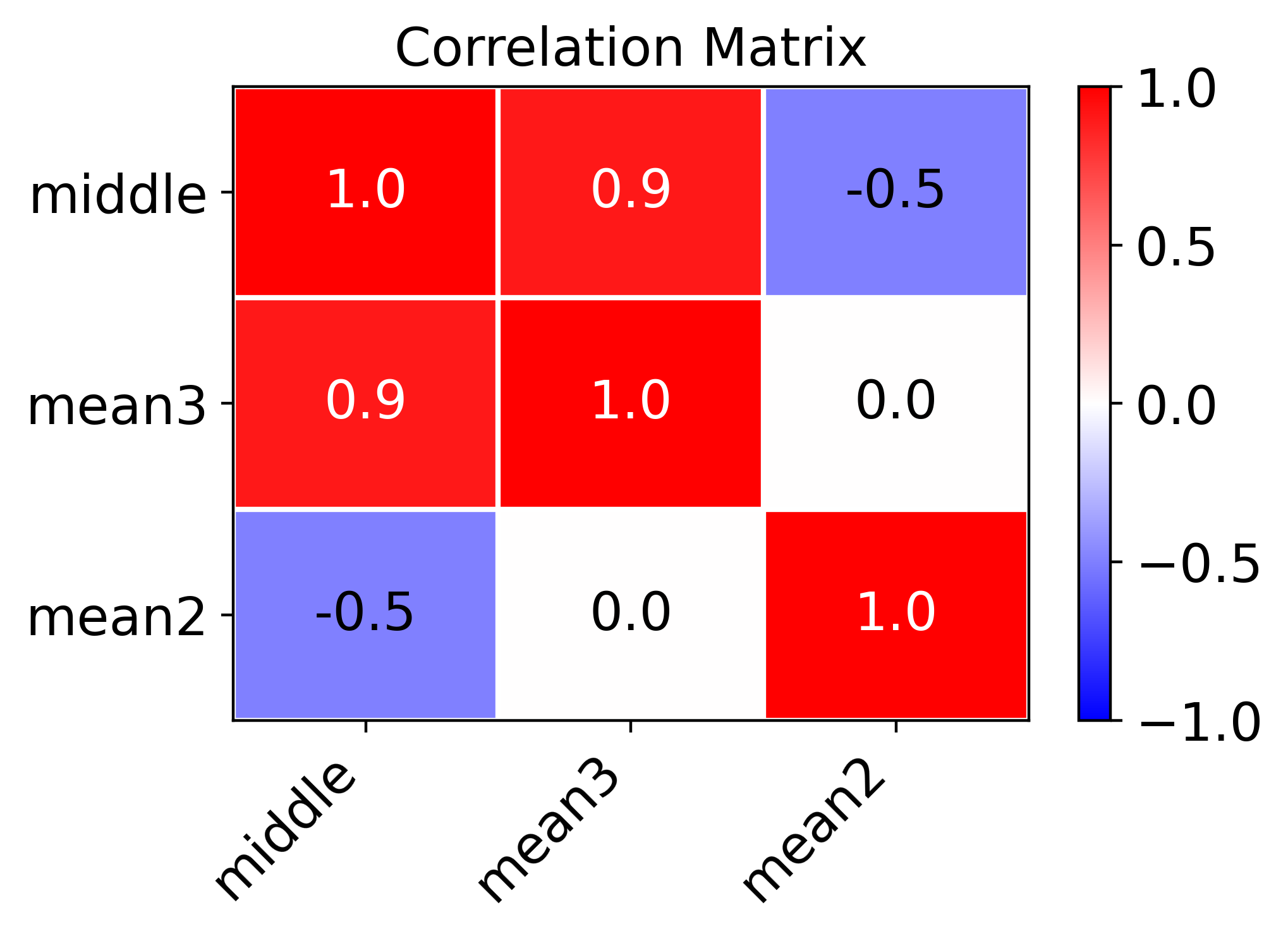}
        \caption{}
        \label{fig:fig11b}
    \end{subfigure}
    \caption{}
        \label{fig:fig11}
\end{figure}

Thermal correlations were also investigated. Figure~\ref{fig:fig11a} demonstrates that, in Case 2, the lateral regions exhibit a strong anti-correlation, where the heating of one region coincides with the cooling of the other, consistent with the Zeroth Law of Thermodynamics. The middle region shows a slight correlation with the lower-temperature zone and a slight anti-correlation with the higher-temperature zone, confirming its role as a heat transporter. Furthermore, Figure~\ref{fig:fig11b} highlights the anti-correlation between the central region's temperature and the average temperature of the lateral ends, further supporting the interpretation of the center as a thermal intermediary. 

\subsection{Correlation of fluctuations and thermal relaxation time}

The concept of thermal relaxation time is closely linked to the dissipation of thermal differences in physical systems. When two regions with different temperatures come into contact, the system tends toward thermal equilibrium according to heat transport laws. Relaxation time quantifies the rate at which this difference diminishes over time.

In statistical mechanics, fluctuations around equilibrium contain information about how the system relaxes. The temporal correlation function of the fluctuations is used to measure the characteristic time of dissipation of the system's memory ~\cite{kubor}. Empirically, this function is obtained by
\begin{equation}
C(t_i) = \langle \delta T(t) \cdot \delta T(t + t_i) \rangle
\end{equation}
in which $\delta T(t) = T(t) - \langle T \rangle$. While thermal fluctuation in statistical mechanics arises directly from the microscopic dynamics of the system and its decay defines a physical relaxation time associated with energy dissipation, the GARCH model describes the statistical evolution of the conditional variance of a time series, capturing the persistence of fluctuations without attributing an underlying physical mechanism to them.

In systems like the one studied in this paper, the shape of the decay of the correlation function of temperature fluctuations can be fitted with exponential functions ~\cite{callen}:
\begin{equation}
C(t) \sim e^{-t/\tau}
\end{equation}
This characteristic time $\tau$ can be interpreted, in our case qualitatively, as the thermal relaxation time—that is, how long it takes for each region to reach thermal equilibrium. Qualitatively, we can also associate it with the thermal conductivity of the wall or the medium.

\begin{figure}[H]
    \centering
    \begin{subfigure}{6cm}
        \centering
        \includegraphics[width=8cm]{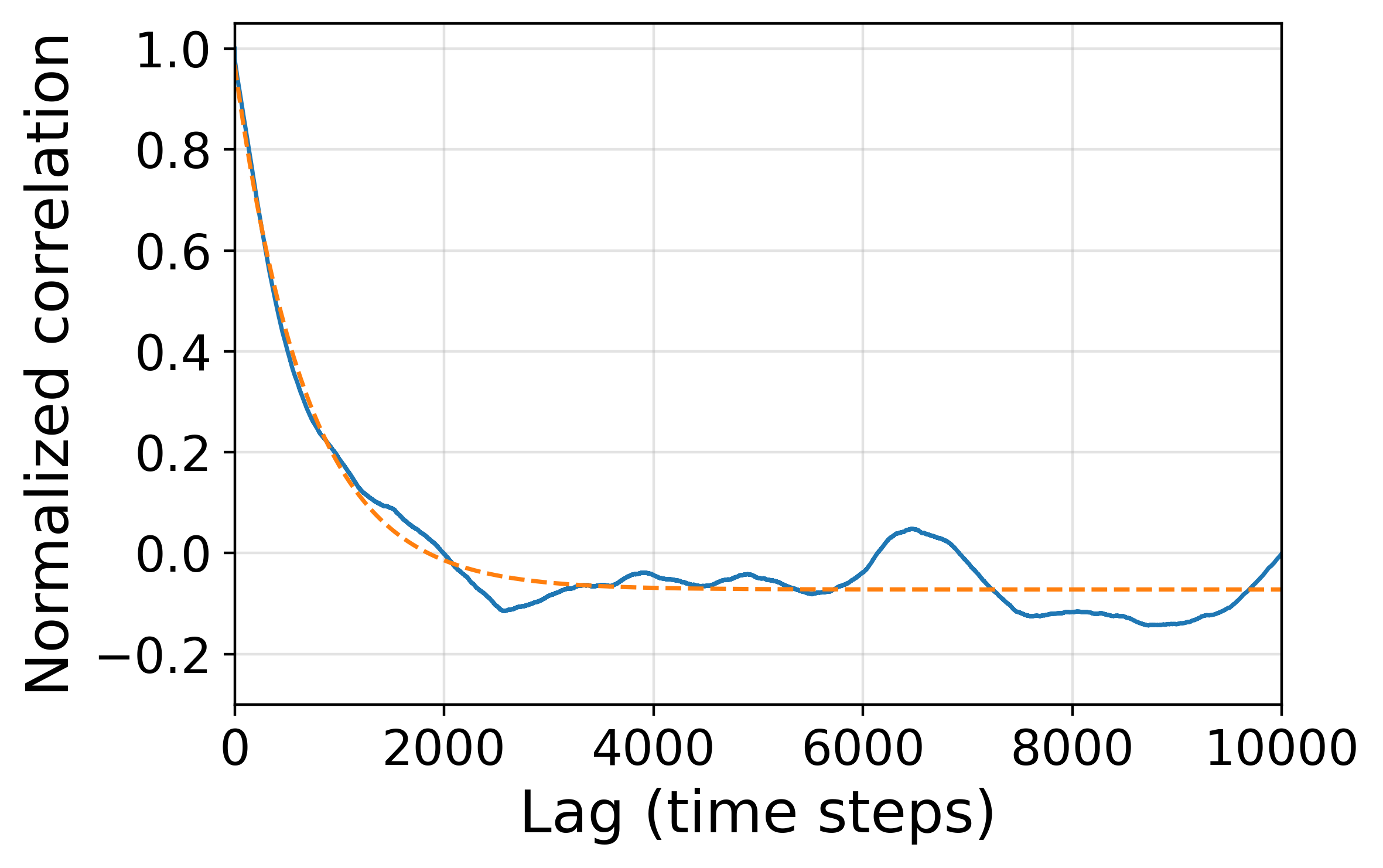}
        \caption{}
        \label{fig12a}
    \end{subfigure}
    \hspace{1.5cm}
    \begin{subfigure}{6cm}
        \centering
        \includegraphics[width=8cm]{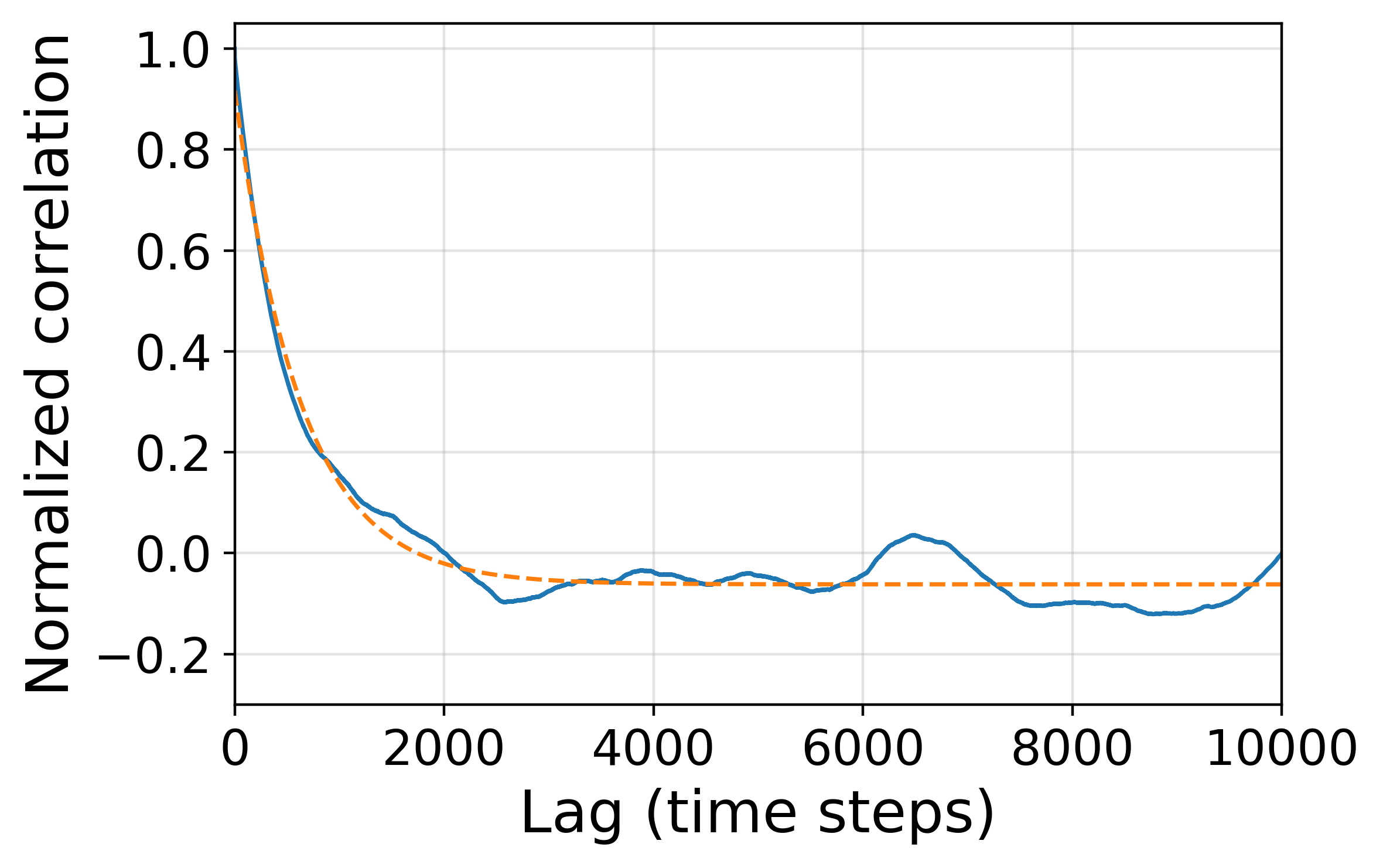}
        \caption{}
        \label{fig12b}
    \end{subfigure}
    \begin{subfigure}{6cm}
        \centering
        \includegraphics[width=8cm]{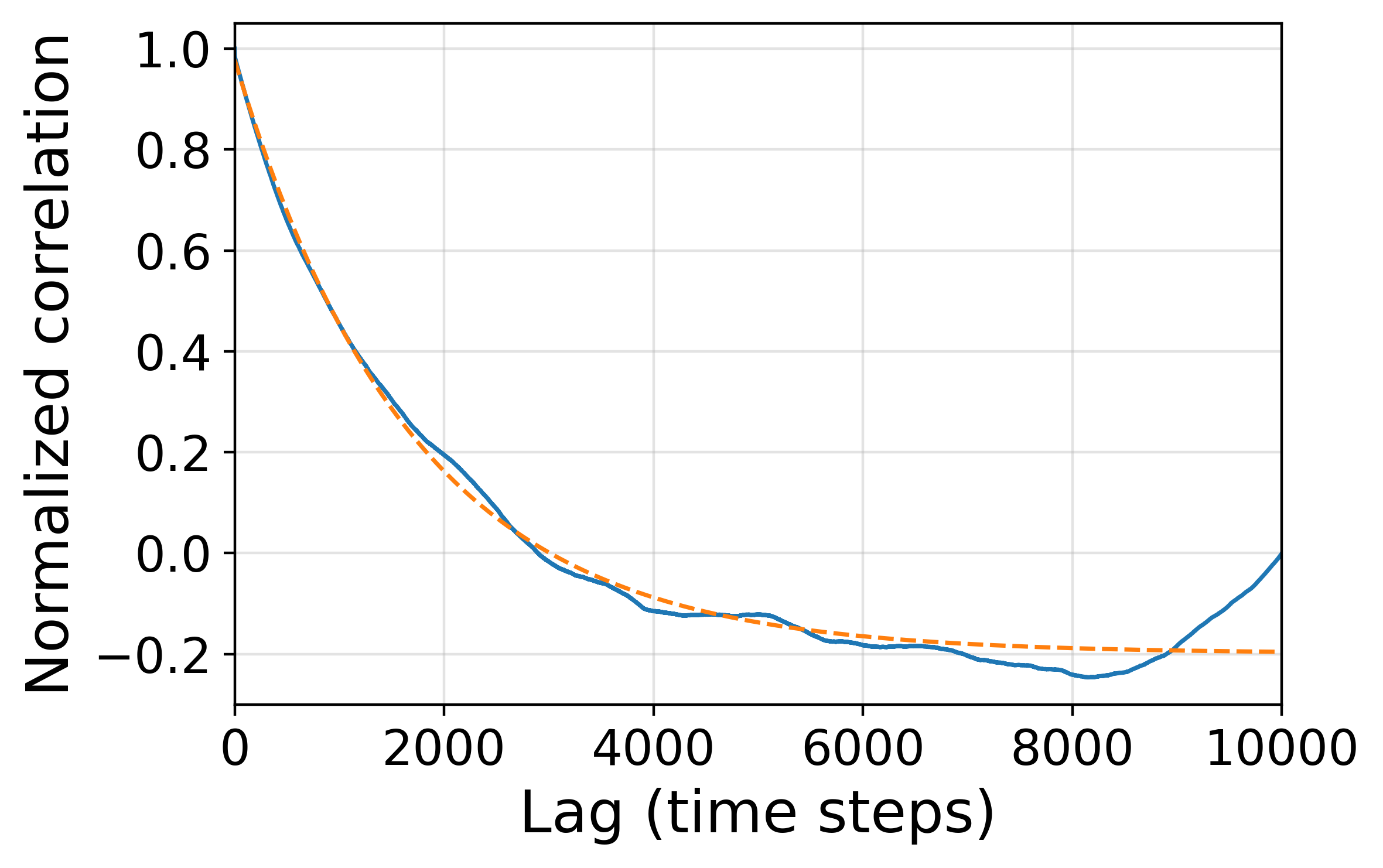}
        \caption{}
        \label{fig12c}
    \end{subfigure}
    \hspace{1.5cm}
    \begin{subfigure}{6cm}
        \centering
        \includegraphics[width=8cm]{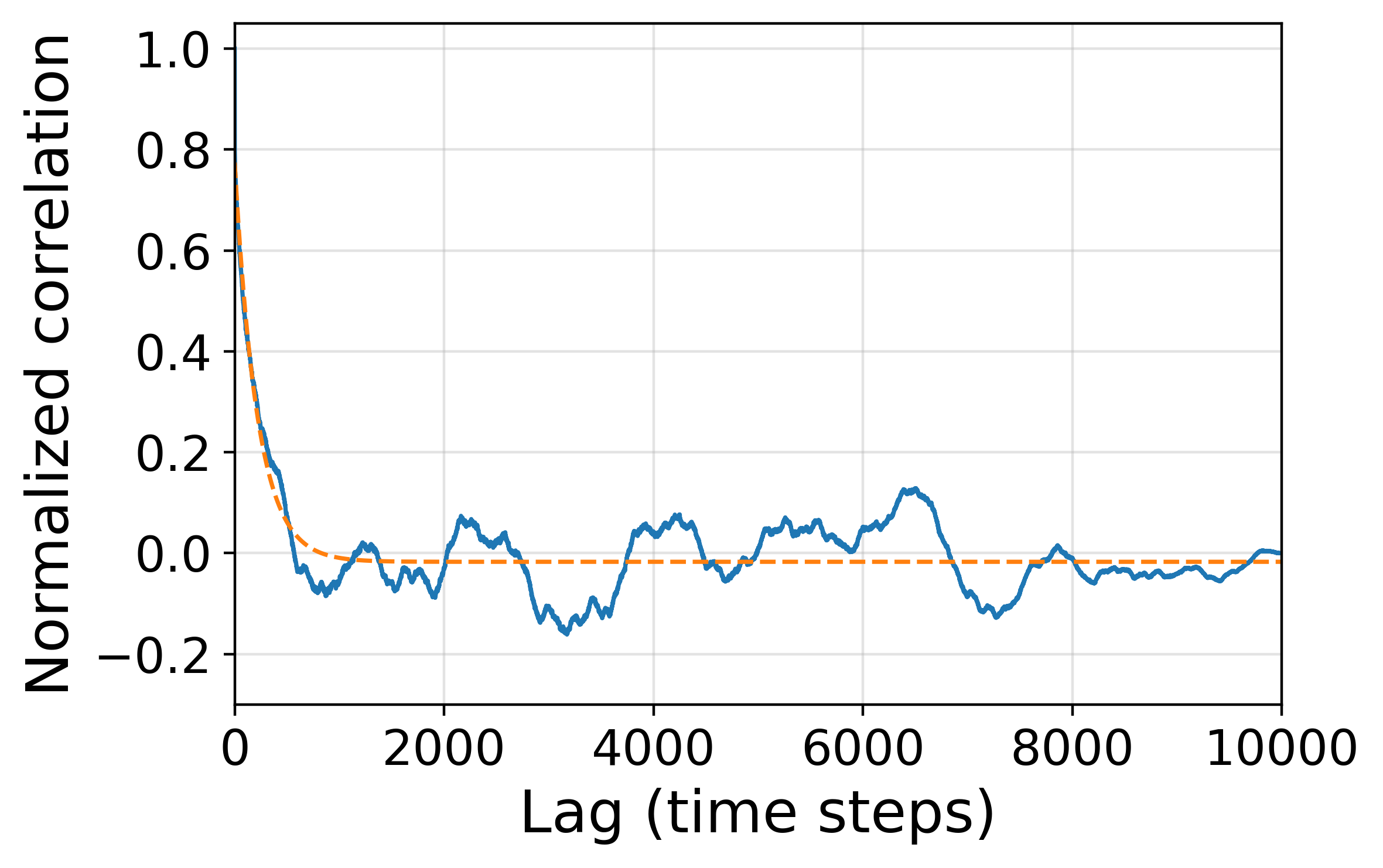}
        \caption{}
        \label{fig12d}
    \end{subfigure}
    \begin{subfigure}{6cm}
        \centering
        \includegraphics[width=8cm]{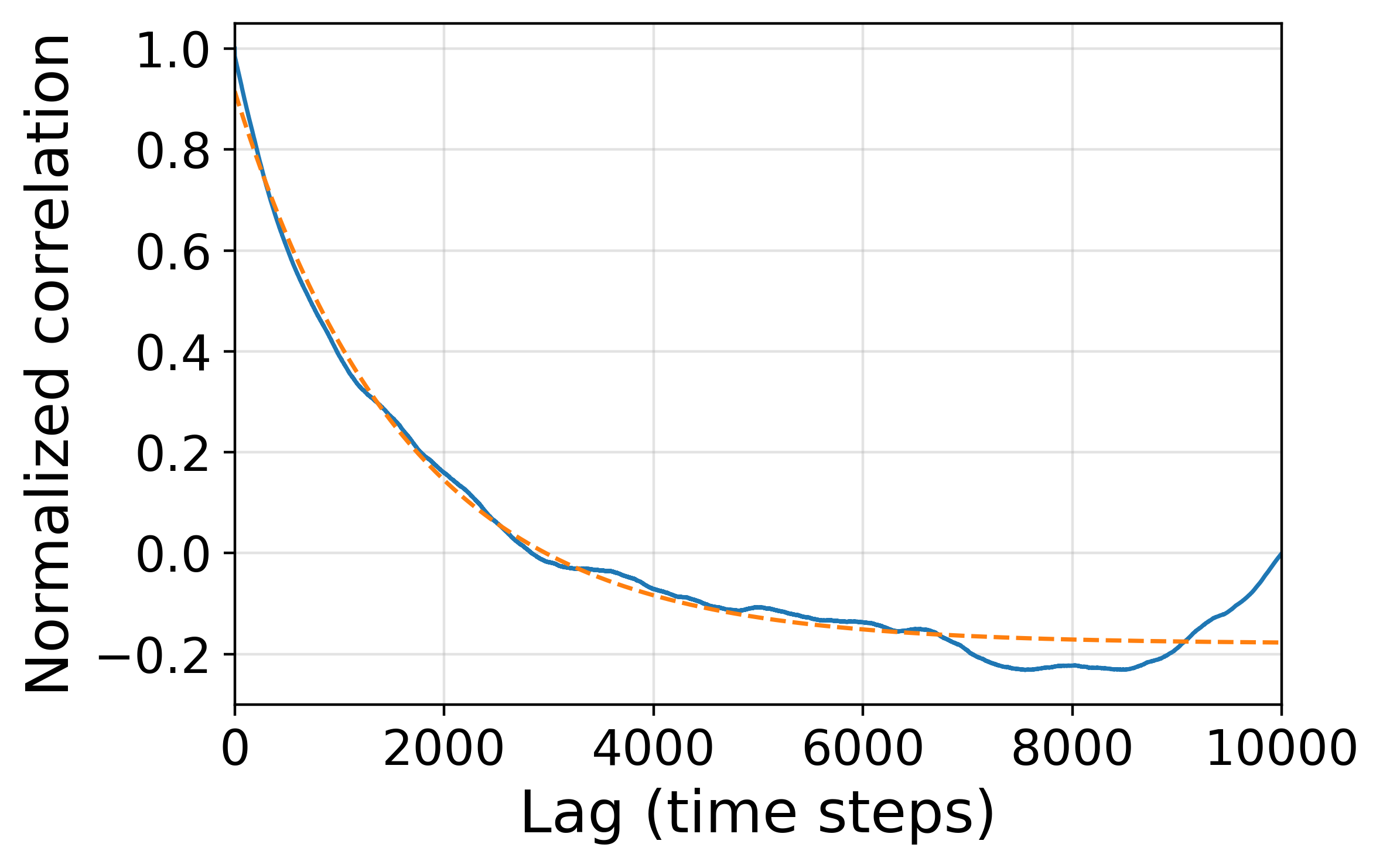}
        \caption{}
        \label{fig12e}
    \end{subfigure}
    
    \caption{Correlation function of temperature as a function of simulation time (solid line) and its exponential fit (dashed line) : Panels (a,b) show the correlations for the left and right regions in a two-region subdivision, whereas panels (c–e) correspond to the left, central, and right regions in a three-region subdivision. Solid curves represent the normalized cross-correlation functions, and dashed curves denote exponential fits from which the characteristic correlation times $\tau$ are extracted.}
        \label{fig12}
\end{figure}

Once the curves for each region of each system studied were obtained, we were able to perform an exponential fit to calculate the relaxation time. As seen in Figure ~\ref{fig12}, the average relaxation time for the two-region system was 662 time steps. For the three-region system, the lateral regions had an average relaxation time of 1669 time steps, and the median region had a value of 216 time steps. Thus, we observed that the system without an intermediate region relaxes 2.5 times faster than the system with an intermediate region. This occurs because, for the lateral regions to reach equilibrium, they need to exchange energy with the atoms in the median region, thus taking longer. Even though the initial temperature of the median region is approximately the equilibrium temperature, the relaxation time, although small, is not zero. This occurs because heat exchanges between the lateral regions produce temperature fluctuations, generating non-zero correlations between them (Figure ~\ref{fig12d}).

\color{black}

\section{Conclusion}

This study demonstrated that the Zeroth Law of Thermodynamics can be more deeply understood by observing the intermediate stages of an ideal gas system as it evolves toward thermal equilibrium. Contrary to the simplified suggestion of the Zeroth Law, the heat conduction process is dynamic and depends on fluctuations, thermal correlations, and the geometric configuration of the system.

The detailed analysis of thermal fluctuations and correlations reveals that while Case 1 reaches equilibrium in a relatively simple and rapid manner, Case 2, which includes a middle region, delays the equilibration process by increasing temperature fluctuations and requiring more time for the system to achieve thermal homogeneity. These findings provide new insights into the mechanisms of heat transport in microscopic systems, offering an expanded perspective on the application of the Zeroth Law of Thermodynamics.

In the three-region configuration, the frequency distribution of temperature fluctuations displays two distinct peaks, identifying both a local and a global maximum. This phenomenon indicates that the system undergoes a multi-stage relaxation process; the local maximum represents a transient 'quasi-equilibrium' achieved between the first two regions, while the global maximum marks the onset of the total system's thermalization. These findings highlight that the path to satisfying the Zeroth Law is non-trivial in non-equilibrium molecular dynamics, as it involves overcoming internal heat conduction bottlenecks that temporarily stabilize inhomogeneous thermal states ~\cite{LiebYngvason1999}.


\end{document}